\documentclass[letterpaper, 10 pt, conference]{ieeeconf}  % Comment this line out if you need a4paper

\IEEEoverridecommandlockouts                              % This command is only needed if
\usepackage{epsfig} % for postscript graphics files
\usepackage{times} % assumes new font selection scheme installed
\usepackage{stmaryrd}
\usepackage{amsfonts}
\usepackage{graphicx,times,amsmath} % Add all your packages here
\usepackage{epsfig} % for postscript graphics files
\usepackage{mathptmx} % assumes new font selection scheme installed
\usepackage{times} % assumes new font selection scheme installed
\usepackage{amsmath,amssymb}
\usepackage{url}
\usepackage{color}
\usepackage{enumerate}

\usepackage[normalem]{ulem}

\newtheorem{theorem}{Theorem}
\newtheorem{lemma}{Lemma}
\newtheorem{remark}{Remark}
\newtheorem{assumption}{Assumption}
\newtheorem{definition}{Definition}

\newtheorem{corollary}{Corollary}

\title{\LARGE \bf
Distributed Consensus of Linear Multi-Agent Systems with Switching Directed Topologies
}

\author{Guanghui Wen$^{1}$ and Valery Ugrinovskii$^{2}$% <-this % stops a space
\thanks{This work was supported by the National Natural Science Foundation
  of China under Grant No. 61304168, the Natural Science
Foundation of Jiangsu Province of China under Grant No. BK20130595 and in
part, by the Australian Research Council through the Discovery Projects
funding scheme (Project DP120102152).}% <-this % stops a space
\thanks{$^{1}$Guanghui Wen is with Department of Mathematics, Southeast University,
Nanjing 210096, China.
        {\tt\small wenguanghui@gmail.com}}%
\thanks{$^{2}$V. Ugrinovskii is with the School of Engineering
and Information Technology, The University of New South Wales Canberra at
the Australian Defence Force Academy, Canberra, ACT 2600, Australia.
{\tt\small v.ougrinovski@adfa.edu.au}}%
}

\begin{document}

\maketitle
\thispagestyle{empty}
\pagestyle{empty}

%%%%%%%%%%%%%%%%%%%%%%%%%%%%%%%%%%%%%%%%%%%%%%%%%%%%%%%%%%%%%%%%%%%%%%%%%%%%%%%%
\begin{abstract}
This paper addresses the distributed consensus
problem for a linear multi-agent system with switching directed communication
topologies. By appropriately introducing a linear transformation,
the consensus problem is equivalently converted to a stabilization problem for a class of switched linear systems. Some sufficient consensus conditions are then
derived by using tools from the matrix theory and stability analysis of
switched systems. It is proved that consensus
 in such a multi-agent system can be ensured if each agent is stabilizable
 and each possible directed topology contains a directed spanning
 tree. Finally, a numerical simulation is given for illustration.
\end{abstract}

%%%%%%%%%%%%%%%%%%%%%%%%%%%%%%%%%%%%%%%%%%%%%%%%%%%%%%%%%%%%%%%%%%%%%%%%%%%%%%%%
\section{INTRODUCTION}
The past few years have witnessed an extensive theoretical development in
the area of cooperative control of multi-agent systems. This area of
control engineering has attracted significant
attention from various scientific communities ranging from mathematics to electronic engineering \cite{SaberFaxMurray2007Survey,RenBeardAtkins2007Survey}.
Much of this attention owes to applications
in the fields of unmanned air vehicles (UAVs), smart grids, large-scale
sensor networks, to name a few
\cite{RenBeardAtkins2007Survey,Ugrinovskii2011Automatica,MeiZhangCaoBook2011}.

\par

One of the central problems in cooperative control of multi-agent systems,
the so called consensus problem \cite{Jadbabaie2003TAC,SaberMurrayTAC2005},
is concerned with the design of local controllers using relative information
to realize a state agreement in the whole group. To date, much progress has
been made in the study of consensus in multi-agent systems. For example,
consensus for first-order multi-agent systems with switching directed
topologies has been investigated in
\cite{RenBeard2005TAC,Moreau2005TAC}. It has
been shown that the state of consensus in such multi-agent systems can
be achieved if and only if the time-varying network topology jointly
contains a directed spanning tree. However, a large class of practical
agents requires higher order dynamical models to describe dynamics of such
systems \cite{RenJIRS2007}. Motivated by this observation, consensus
problems for second-order multi-agent systems and integrator-type
higher-order multi-agent systems have been studied in
\cite{RenAtkinsRNC2007,WenDuanYuChenRNC2012,WielandKimFrankIJSS2011}. % In
% view of the agents' dynamics, multi-agent systems with general linear node
% dynamics (linear multi-agent systems) include those with
%  integrator-type dynamics (of any order) as special cases.
 More recently, the consensus problem for general linear multi-agent systems has
received a lot of  attention
 \cite{LiDuanChenHuangTCASI2010}-\cite{WenDuanRenChenRNC}. Specifically,
the consensus problem for linear multi-agent systems under a fixed directed
 communication topology has been addressed in
 \cite{LiDuanChenHuangTCASI2010}, \cite{ZhangFrankDas2011TAC}. In
 \cite{TrentelmanTakabaMonshizadehIEEETAC2013}, the robust consensus of linear
 multi-agent systems with additive perturbations of the transfer matrices
 of the nominal dynamics was studied. In~\cite{Ugrinovskii2011Automatica}
 and a number of subsequent papers, the robust consensus  was analyzed from
 the view point of the $H_\infty$ control theory.
Among other relevant references, we mention \cite{SuHuangTAC2012} and
\cite{WuQinYuAllgowerCDC2013} where, while assuming that the open loop
systems are Lyapunov stable, the consensus problems for linear multi-agent
systems with undirected and directed switching topologies have been
investigated, respectively. In the situation where the multi-agent
system is equipped with a leader and the topology of the system
belongs to the class of switched directed topologies, the consensus
tracking problem has been studied in
\cite{WenHuYuChenCaoSCL2013,WenDuanRenChenRNC}. One  feature of the results in these references
 is that the open loop agents' systems were not
required to be Lyapunov stable. Note that the presence of the leader in
the multi-agent systems considered in these reference facilitated the
derivation and a direct
analysis of the consensus error system. However, when the open loop systems are
not Lyapunov stable and/or there is no designated leader in the group, the
consensus problem for linear multi-agent systems with switching directed
topologies remains challenging.

 \par

Motivated by the above discussion, this paper aims to study the consensus
problem for linear multi-agent systems with switching directed
topologies. Several aspects of our study are worth mentioning. Firstly, we
dispense with some of the assumptions in the existing work. E.g., the open
loop dynamics of the agents are not required to be Lyapunov stable
in the present paper. Furthermore, the multi-agent systems under
consideration are not required to have a leader. Compared to
consensus problems for linear multi-agent systems that have a
designated leader, the point of difference here concerns the assumption on
the communication topology of the system. In the previous work on
the linear leader-tracking multi-agent consensus such as, e.g.,
\cite{WenHuYuChenCaoSCL2013}, each possible augmented system graph was
required to contain a directed spanning tree rooted at the leader. Compared
to that work, the switching topologies in the present paper are allowed to
have spanning trees rooted at different nodes. This is a significant
relaxation of the existing conditions since it enables the system to be
reconfigured if necessary (e.g., to allow different nodes to serve as the
formation leader). This also has a potential to make the system more
reliable.

%\uuline{it has been shown in
%\cite{WenHuYuChenCaoSCL2013} that consensus tracking problem for such a
%multi-agent system can be achieved if the}. \textbf{Unfinished sentence?
%Also, you might want to mention that the paper provides conditions for the
%design of the switching sequence (i.e, selection of the dwell time) and
%that switching graphs are allowed to have spanning trees with different
%rooted nodes.}
The analysis of leaderless tracking consensus in this paper proceeds as
follows. By introducing a linear transformation, the consensus problem for
the leaderless multi-agent system under consideration is first converted to
the global uniform asymptotic stability analysis problem for a class of
switched linear systems with zero equilibrium. It is worth noting that the
transformation technique employed in this paper is different from those
used in \cite{MinyiHuangTAC,WenwuYuSMCB} where the consensus problem was
converted to the global asymptotic stability problem over a time-varying
manifold.  Then, by constructing multiple Lyapunov functions,
sufficient conditions are derived for the system under consideration to
achieve consensus. Similar to \cite{RenBeard2008Book}, our
results impose a condition on the minimum dwell time of the network, i.e.,
the minimum time the system must spend in a particular
configuration. Recall that the conditions in \cite{RenBeard2008Book}
provide a lower bound on the dwell time which ensures that consensus in
a multi-agent system of coupled double integrators with switching topology
is achieved if each possible topology contains a directed spanning
tree. However, the analysis method given in \cite{RenBeard2008Book} is
inapplicable to general linear multi-agent
systems since one can not directly calculate the divergence behavior of
general linear multi-agent systems.

\par

The remainder of this paper is structured as follows. In
Section II, some preliminaries from the graph theory and
the problem formulation are given. In Section III, the main
theoretical results are presented. A numerical example is provided in Section
IV for illustration. Finally, Section V concludes the paper.

\par
Throughout the paper, let $\mathbf{R}^{n\times n}$ and $\mathbf{N}$ be the sets of $n\times n$ real
matrix space and positive natural numbers, respectively. The superscript
$T$ denotes the transpose of a
matrix. Unless explicitly stated otherwise,
  all matrices are assumed to have
compatible dimensions. The matrix inequality $A>B$ means that both $A$ and $B$
are symmetric matrices and that $A-B$ is positive-definite. For a complex
number $\theta$, $\mathbf{Re}(\theta)$
  denotes its real part.
Let $I_{n}$ be the $n$-dimensional identity matrix. The notation
$\mathbf{1}_n$ and $\mathbf{0}_{n}$ refers to the $n$-dimensional column
vectors with all entries equal to $1$ and $0$, respectively.
The symbols $\otimes$ and $\|\cdot\|$ denote, respectively, the Kronecker
product and the Euclidean norm.

\section{Preliminaries and Problem Formulation}

In this section, some preliminaries from the algebraic graph theory and the
problem formulation are given.
\subsection{Preliminaries}
The communication topology of the considered multi-agent system will be
described by a switching directed graph.
Let $\mathcal{G}(\mathcal{V},\mathcal{E},\mathcal {A})$ be a directed graph with the set of nodes $\mathcal{V} = \left\{ {1,2, \cdots ,N} \right\}$, the set of
directed edges $\mathcal{E}\subseteq \left\{(i,j),\,i,j\in
  \mathcal{V}\right\}$  and a weighted adjacency matrix
$\mathcal{A}=[a_{ij}]_{N\times N}$ with non-negative elements. The pair
$(i,j)$ denotes the edge
of the graph $\mathcal{G}(\mathcal{V},\mathcal{E},\mathcal {A})$
originating at node $j$ and ending at node $i$.
The adjacency matrix $\mathcal{A} = [{a_{ij}}]_{N \times N}$  of a directed
graph $\mathcal{G}(\mathcal{V},\mathcal{E},\mathcal {A})$ is defined by
${a_{ii}} = 0$  for $i = 1,2, \cdots ,N$,  and ${a_{ij}}> 0$ for $(i,j) \in
\mathcal{E}$ but $0$ otherwise. The Laplacian matrix $\mathcal{L} =
[{l_{ij}}]_{N \times N}$ is defined as ${l_{ij}} = -a_{ij},$ $i \ne j,$
and  $l_{ii}=\sum\nolimits_{k = 1}^N {{a_{ik}}} $  for  $i = 1,2, \cdots
,N$. In the sequel, we will
use a shorthand notation $\mathcal{G}(\mathcal{A})$ in lieu of
$\mathcal{G}(\mathcal{V},\mathcal{E},\mathcal {A})$ if no confusion
arises.  A path in
$\mathcal{G}(\mathcal{A})$ from node ${i_1}$ to node ${i_s}$ is a sequence
of ordered edges of the form $({i_{k+1}},\;{i_k}),$ $k = 1,2, \cdots ,s-1$.
A directed graph is strongly connected if and only if there exists a path
between every pair of distinct nodes.  A directed
graph is said to contain a spanning tree if there
exists a node called the root node such
that there exists a
directed path from this root node to every
other node.

\par
In this paper, the communication topology of
the multi-agent systems under
consideration is assumed to be dynamically
 switching over a graph set $\widehat{\mathcal{G}}$, where $\widehat{\mathcal{G}}=\{\mathcal{G}(\mathcal{A}^{(1)}),\mathcal{G}(\mathcal{A}^{(2)}),
 \cdots,\mathcal{G}(\mathcal{A}^{(p)})\}$, $p\geq 1$, denotes the set of
 all possible directed topologies. Note that it is assumed in the present
 work that the graphs $\mathcal{G}(\mathcal{A}^{(i)}$, $1\leq i \leq p$, share
 the same node set $\mathcal{V}$.

\subsection{Problem formulation}
Consider a multi-agent systems of $N$ agents, indexed by $1, 2, \cdots, N$. The dynamics
of agent $i$ are described by
\begin{equation}\label{Mode}
\dot{x}_i(t) = A{x_i}(t) + B{u_i}(t),
\end{equation}
where $x_{i}(t)\in \mathbf{R}^{n}$ is the state, ${u_i}(t) \in {\mathbf{R}^m}$ is the control input,
 $A$ and $B$ are constant real matrices with compatible dimensions. It is assumed that matrix pair $(A,B)$
is stabilizable.

Within the context of multi-agent systems, only relative information among neighboring agents can be used for coordination. For each agent $i$, the following distributed consensus protocol is proposed
\begin{eqnarray} \label{CommunicationProtocol}
  u_{i}(t)=\alpha K\sum_{j=1}^{N}\!a_{ij}(t)\!\left[x_{j}(t){-}x_{i}(t)\right],\;\;i=1,2,\cdots,N,
   \end{eqnarray}
where $\alpha>0$ represents the coupling strength, $K \in
\mathbf{R}^{m\times n}$ is the feedback gain matrix to be designed, and $\mathcal{A}(t)=\big[a_{ij}(t)\big]_{N\times N}$ is the
adjacency matrix of graph $\mathcal{G}(\mathcal {A}(t))$. Here,
$\mathcal{G}(\mathcal {A}(t))$ describes the underlying communication
topology of the agents at time $t$. Then, it follows from (\ref{Mode}) and
(\ref{CommunicationProtocol}) that
 \begin{equation}\label{ClosedLoopMAs}
\dot {x}_{i}(t) =Ax_{i}(t)+\alpha BK\sum_{j=1}^{N}{a}_{ij}(t)[x_{j}(t)-x_{i}(t)],
\end{equation}
where $i=1,2,\cdots,N$.
 \par

\begin{definition}\label{ConsensusDefinition}
 The consensus problem of multi-agent system (\ref{Mode}) is solved by
 protocol (\ref{CommunicationProtocol}) if, for any initial conditions, the
 states of the close-loop system (\ref{ClosedLoopMAs}) satisfy
\begin{equation}
{\lim _{t \to \infty }}\left\| {{x _i}(t) - {x _j}(t)} \right\| = 0,\;\;\;\forall \;i,\,j = 1,2, \cdots, N.
\end{equation}
\end{definition}
\par
Suppose that $\mathcal{G}(\mathcal {A}(t))\in \widehat{\mathcal{G}}$ for all $t$. To describe the time-varying property of communication topology, assume that there exists
an infinite sequence of non-overlapping time intervals $[t_{k},\,t_{k+1})$,
$k\in \mathbf{N}$, with $t_{1}=0$, $\tau_{1}>t_{k+1}-t_{k}\geq \tau_{0}>0$,
over which the communication topology is fixed. Here, $\tau_{1}>\tau_{0}>0$
and  $\tau_{0}$ is called the dwell time.
The introduction of the switching signal
$\sigma(t):\,[0,\,+\infty)\rightarrow \{1,2,\cdots,p\}$ makes the
communication topology of multi-agent system (\ref{Mode}) well defined at
every time $t$, where $t\geq 0$. For notational convenience, we will
describe this communication topology using the time-varying graph
$\mathcal{G}(\mathcal{A}^{\sigma(t)})$.

\par
Take $x(t)=\left[x_{1}^{T}(t),x_{2}^{T}(t),\cdots,x_{N}^{T}(t)\right]^{T}$, it thus follows from (\ref{ClosedLoopMAs}) that
\begin{equation}\label{Closed-loop-vector}
{\dot {x}}(t) = \left[(I_{N}\otimes A) - \alpha \left(\mathcal{L}^{(\sigma(t))}\otimes BK \right)\right] x(t),
\end{equation}
where $\mathcal{L}^{(\sigma(t))}$ is the Laplacian matrix of communication topology $\mathcal{G}(\mathcal{A}^{(\sigma(t))})$.

Before concluding this section, the following assumption is presented which
will be used in the derivation of  the main results.
\begin{assumption}\label{AssumptionTopology}
For each $i\in \{1,2,\cdots,p\}$, the graph $\mathcal{G}(\mathcal{A}^{(i)})$ contains a directed spanning tree.
\end{assumption}

\begin{remark}
Note that we will not assume in the sequel that the directed spanning trees
within the graphs $\mathcal{G}(\mathcal{A}^{(i)})$, $i= 1,2,\cdots,p$, share a
common root node, though such an assumption is very common in the existing
related literature
\cite{WenHuYuChenCaoSCL2013,WenYuCaoHuChenASCC2013,GaoZhuChenZhangMPE2013}.
Certainly, Assumption \ref{AssumptionTopology} holds in the special case
considered in the above references, where
each possible topology $\mathcal{G}(\mathcal{A}^{(i)})$, $i\in
\{1,2,\cdots,p\}$, contains a directed spanning tree, and all these trees
are rooted at the same node. Furthermore,
Assumption \ref{AssumptionTopology} will hold if each possible topology is
strongly connected.
\end{remark}
\par
Note that all signals considered in this paper are assumed to be
differentiable on the right. Furthermore, for any given  $x(0)\in
\mathbf{R}^{Nn}$, the switched systems (\ref{Closed-loop-vector}) are
assumed to have a unique and absolutely
continuous solution $x(t;x(0))$ in the sense of Carath\'{e}odory
\cite{DianelBook2003}.

\begin{lemma}[\cite{HornMatrix}]\label{lemmaInequality}
For any given $\zeta$, $\eta\in \mathbf{R}^{n}$, and matrices $P>0$, $D$ and $S$ of appropriate
dimensions, one has
\begin{equation}
2\zeta^{T}DS \eta\leq \zeta^{T}DPD^{T}\zeta+\eta^{T}S^{T}P^{-1}S\eta.
\end{equation}
\end{lemma}

\section{Main Results}
In this section, the main theoretical results are presented and analyzed.
\par
\subsection{Consensus of Linear Multi-Agent Systems with Switching Directed Topology}
Let $e(t)=[e_{1}^{T}(t),e_{2}^{T}(t),\cdots,e_{N-1}^{T}(t)]^{T}$, with
$e_{i}(t)=x_{i}(t)-x_{N}(t)$, for $i=1,2,\cdots,N-1$.
Here, $e(t)$ represents the group
\emph{disagreement vector} \cite{SaberMurrayTAC2005}.
It can then be
obtained that $e(t)=(\Xi\otimes I_{n})x(t)$, where
$\Xi=\left[I_{N-1},\,-\mathbf{1}_{N-1}\right]\in \mathbf{R}^{(N-1)\times
  N}$. Using this notation, it can thus be obtained from
(\ref{Closed-loop-vector}) that
\begin{eqnarray}\label{ClosdLoopConsensusErrorReducedVersion}
\dot{e}(t)=(I_{N-1}\otimes A)e(t)-\alpha\left(\Xi\mathcal{L}^{(\sigma(t))} \otimes BK\right)x(t).
\end{eqnarray}
Noticing that
$
\left(\mathcal{L}^{(\sigma(t))} \otimes BK\right)(\mathbf{1}_{N}\otimes I_{n})x_{N}(t)=\mathbf{0},
$
one has that
\begin{eqnarray}\label{PropertyLaplacianMatrix}
\left(\Xi\mathcal{L}^{(\sigma(t))} \otimes BK\right)x(t)=\left(\Xi\mathcal{L}^{(\sigma(t))}\Pi \otimes BK\right)e(t),
\end{eqnarray}
where $\Pi=\left[\begin{array}{c}
I_{N-1}\\
\mathbf{0}_{N-1}^{T}
\end{array}\right]\in \mathbf{R}^{N\times (N-1)}$. Substituting (\ref{PropertyLaplacianMatrix}) into
(\ref{ClosdLoopConsensusErrorReducedVersion}) gives that
\begin{eqnarray}\label{ClosdLoopConsensusErrorRewritten}
\dot{e}(t)=\left[I_{N-1}\otimes A-\alpha\left(\Xi\mathcal{L}^{(\sigma(t))}\Pi \otimes BK\right)\right]e(t).
\end{eqnarray}
Obviously, $\mathbf{0}$ is the equilibrium point of the switched system
(\ref{ClosdLoopConsensusErrorRewritten}). Furthermore, by Definition
\ref{ConsensusDefinition}, the multi-agent system (\ref{ClosedLoopMAs})
achieves consensus if and only if the zero equilibrium point of the switched
system (\ref{ClosdLoopConsensusErrorRewritten}) is globally
attractive. Thus,  to show that
the multi-agent system (\ref{ClosedLoopMAs}) achieves consensus it is
sufficient to establish that the zero
equilibrium point of the switched system
(\ref{ClosdLoopConsensusErrorRewritten}) is globally asymptotically
stable.

According to Assumption \ref{AssumptionTopology}, it can be
obtained from Theorem 2.8 in \cite{RenBeard2008Book} that, for each $i\in
\{1,2,\cdots,p\}$ and an arbitrarily given $\alpha>0$, the linear
time-invariant system
\begin{eqnarray}
\dot{\zeta}(t)=-\alpha\left(\Xi\mathcal{L}^{(i)}\Pi \otimes I_{n}\right)\zeta(t),
\end{eqnarray}
is globally asymptotically stable about its zero equilibrium point, where
$\zeta(t)\in \mathbf{R}^{(N-1)n}$. This implies that for each
$i\in \{1,2,\cdots,p\}$, the $(N-1)\times (N-1)$ matrix
$\Xi\mathcal{L}^{(i)}\Pi$ is antistable. For notational convenience, let
$\widehat{\mathcal{L}}^{(i)}=\Xi\mathcal{L}^{(i)}\Pi$,
$i=1,2,\cdots,N-1$. Choose a positive scalar
$c_{i}<\lambda_{\mathrm{min}}^{(i)}$, where  $\lambda_{\mathrm{min}}^{(i)}=\mathrm{min}_{j=1,2,\cdots,N-1}\mathbf{Re}(\lambda_{j}(\widehat{\mathcal{L}}^{(i)}))$, and
$\lambda_{j}(\widehat{\mathcal{L}}^{(i)})$, $j=1,2,\cdots, N-1$, are the
eigenvalues of $\widehat{\mathcal{L}}^{(i)}$.
Then it is easy to verify that there
exists a positive definite matrix $Q^{(i)}$ such that
\begin{eqnarray}\label{LMILaplacianMatrix}
\left(\widehat{\mathcal{L}}^{(i)}\right)^{T}Q^{(i)}+Q^{(i)}\widehat{\mathcal{L}}^{(i)}>2c_{i} Q^{(i)},
\end{eqnarray}
\par
\begin{remark}
By introducing a linear transformation, the
  consensus problem for the multi-agent system
(\ref{ClosedLoopMAs}) is transformed into the
problem of stabilizing globally
the switched linear system
(\ref{ClosdLoopConsensusErrorRewritten}). Note that the dynamics of the group
disagreement vector $e$ can be directly obtained when there is a common leader
in the multi-agent systems or each possible topology is strongly connected
and balanced \cite{WenHuYuChenCaoSCL2013,WenYuCaoHuChenASCC2013}. It is
also worth noting that the
transformation matrix $\Xi$ in (\ref{ClosdLoopConsensusErrorRewritten}) is
not unique \cite{WenwuCaoLuSIAM2007,SunWangXieTAC2008}.
\end{remark}
\par
% Before moving on, the following algorithm is given
% to design the feedback gain matrix and coupling strength of protocol (\ref{CommunicationProtocol}) for achieving
% consensus tracking in the closed-loop system (\ref{ClosedLoopMAs}).

The following theorem presents
the design of a feedback gain matrix and
the coupling strength
for the protocol
(\ref{CommunicationProtocol}) to achieve
consensus tracking by the closed-loop system (\ref{ClosedLoopMAs}).
%
%
%\begin{algorithm}\label{AlgorithmOne}
%Suppose that the matrix pair $(A, B)$ is stabilizable
%and Assumption \ref{AssumptionTopology} holds. Then, the consensus protocol (\ref{CommunicationProtocol}) can be
%designed according to the following procedures:
%\begin{itemize}
%  \item [i)] Set $\beta>0$, solving the following linear matrix inequality (LMI):
%  \begin{eqnarray}\label{LMISolvingFeedbackgainMatrix}
%  AP+PA^{T}-BB^{T}+\beta P<0,
%  \end{eqnarray}
%  yields a matrix $P>0$.
%  Then, take $K = (1/2)B^{T}P^{-1}$
%  \item [ii)] Select the coupling strength $\alpha>2/c_{0}$, where $c_{0}=\mathrm{min}_{i=1,2,\cdots,N-1}c_{i}$, and $c_{i}$, $i=1,2,\cdots,N-1$, are given in (\ref{LMILaplacianMatrix}).
%\end{itemize}
%\end{algorithm}
%
%\textbf{I am not sure why this is an algorithm. This looks like a list of
%  conditions for Theorem 1. Wouldn't it be easier to simply say: \emph{Suppose
%  $\beta>0$ is such that (\ref{LMISolvingFeedbackgainMatrix}) has a
%  positive definite symmetric solution, and  $\alpha>2/c_{0}$, where
%  $c_{0}=\mathrm{min}_{i=1,2,\cdots,N-1}c_{i}$, and $c_{i}$,
%  $i=1,2,\cdots,N-1$, are given in (\ref{LMILaplacianMatrix}). Then the
%  multi-agent system (\ref{ClosedLoopMAs}) with protocol \ldots achieves
%  consensus if \ldots }.}
%
This theorem summarizes the main theoretical results of this paper.
\begin{theorem}\label{MainTheoremOne}
Suppose that Assumption \ref{AssumptionTopology} holds and there exists a
$\beta>0$ such that the following linear matrix inequality (LMI)
\begin{eqnarray}\label{LMISolvingFeedbackgainMatrix}
\label{LMISolvingFeedbackgainMatrixCorollary}
  AP+PA^{T}-BB^{T}+\beta P<0,
  \end{eqnarray}
has a feasible solution $P>0$. Then, the multi-agent system
(\ref{ClosedLoopMAs}) with $K=(1/2)B^{T}P^{-1}$ achieves consensus if the
following conditions hold:
\begin{enumerate}[(i)]
\item
The coupling strength $\alpha$ satisfies the condition $\alpha>2/c_{0}$ where
$c_{0}=\mathrm{min}_{i=1,2,\cdots,N-1}c_{i}$, and $c_{i}$,
$i=1,2,\cdots,N-1$, are given in (\ref{LMILaplacianMatrix}); and
\item
For some $\kappa_0>0$, the switching interconnection graph
$\mathcal{G}(\mathcal{A}^{\sigma(t)})$
satisfies the following condition
\begin{eqnarray}\label{TheoremCondiiton}
\beta (t_{k+1}-t_{k}) - \mathrm{ln}\overline{\lambda}_{\mathrm{max}}^{(k)}>
\kappa_0,
\end{eqnarray}
% there exists a positive constant $\kappa_{0}>0$ such that
% \begin{eqnarray}\label{TheoremCondiiton}
% \beta (t_{k+1}-t_{k})>\mathrm{ln}\overline{\lambda}_{\mathrm{max}}^{(k)}+\kappa_{0},
% \end{eqnarray}
where $\overline{\lambda}_{\mathrm{max}}^{(k)}$ is the largest eigenvalue
of $\left(Q^{(\sigma(t_{k}))}\right)^{-1}Q^{(\sigma(t_{k+1}))}$, $k\in
\mathbf{N}$.
\end{enumerate}
\end{theorem}

\begin{remark}
It can be seen that the existence
of the protocol (\ref{CommunicationProtocol}) depends on the feasibility of
the LMI (\ref{LMISolvingFeedbackgainMatrix}).
In the case where the pair $(A,B)$ is controllable, the LMI
(\ref{LMISolvingFeedbackgainMatrix}) is feasible for any given $\beta>0$.
In the case of stabilizable  but not completely controllable $(A,B)$,
denote by $\widetilde{\lambda}_{i}$, $i=1,2,\cdots,s$, all the
uncontrollable modes of $(A,B)$. Then the LMI
(\ref{LMISolvingFeedbackgainMatrix}) is feasible if and only if
$\beta<\mathrm{min}_{i=1,2,\cdots,s}\mathbf{Re}(-\widetilde{\lambda}_{i})$.
\end{remark}

From Theorem~\ref{MainTheoremOne}, the following corollary follows which
sets a sufficient condition on the communication topology, in terms of its
dwell time, for the system under consideration to achieve consensus.

\begin{corollary}\label{CorollaryOne}
Suppose that Assumption \ref{AssumptionTopology} holds and there exists a
$\beta>0$ such that the LMI (\ref{LMISolvingFeedbackgainMatrix})
% \begin{eqnarray}\label{LMISolvingFeedbackgainMatrixCorollary}
%   AP+PA^{T}-BB^{T}+\beta P<0,
%   \end{eqnarray}
has a feasible solution $P>0$. Then,
the multi-agent system
(\ref{ClosedLoopMAs}) with $K=(1/2)B^{T}P^{-1}$ achieves consensus if the
coupling strength $\alpha$ satisfies the condition $\alpha>2/c_{0}$ where
$c_{0}$ was defined in Theorem~\ref{MainTheoremOne}, and the dwell time of
the switching communication graph $\mathcal{G}(\mathcal{A}^{\sigma(t)})$
satisfies the following condition
\begin{eqnarray}\label{TheoremCondiitonCorollary}
\tau_{0} >\frac{\mathrm{ln}\overline{\lambda}_{\mathrm{max}}}{\beta},
\end{eqnarray}
where $\overline{\lambda}_{\mathrm{max}}=\mathrm{max}_{i,j=1,2,\cdots,p, \,i\neq j}\widehat{\lambda}_{i,j}$, where $\widehat{\lambda}_{i,j}$ is the largest eigenvalue of $\left(Q^{(i)}\right)^{-1}Q^{(j)}$.
\end{corollary}
% \begin{proof}
% According to (\ref{TheoremCondiitonCorollary}), one knows that there exist a positive scalar $\kappa_{1}>0$ such that $\tau_{0} >\frac{\mathrm{ln}\overline{\lambda}_{\mathrm{max}}}{\beta}+\kappa_{1}$, i.e., $\beta\tau_{0} >\mathrm{ln}\overline{\lambda}_{\mathrm{max}}+\beta\kappa_{1}$. Construct the following multiple Lyapunov functions
% for the switched system (\ref{ClosdLoopConsensusErrorRewritten}):
% \begin{equation}\label{LyapunovFunction}
% V(t)=e^{T}(t)\left(Q^{(\sigma(t))}\otimes P^{-1}\right)e(t),
% \end{equation}
% where $Q^{(\sigma(t))}\in \{Q^{(1)},Q^{(2)},\cdots,Q^{(N-1)}\}$, $Q^{(i)}$, $i=1,2,\cdots,N-1$, are defined in (\ref{LMILaplacianMatrix}), $P>0$  is a
% solution of the LMI (\ref{LMISolvingFeedbackgainMatrixCorollary}). The
% statement of the corollary can then be proved using the reasoning similar
% to that used in the proof of Theorem~\ref{MainTheoremOne}.
% \end{proof}

%\textbf{It is important to explain what is the advantage and disadvantage
%  of using this corollary. For instance, does using
%  $\left(Q^{(i)}\right)^{-1}Q^{(j)}$ instead of
%  $\left(Q^{(\sigma(t_{k}))}\right)^{-1}Q^{(\sigma(t_{k+1}))}$ for the
%  design of dwell time bring significant conservatism? Note that
%  $\left(Q^{(i)}\right)^{-1}Q^{(j)}$ involves the matrices which correspond
%  to time instances which are not time-ordered.}
\begin{remark}
Compared to Theorem \ref{MainTheoremOne}, the consensus conditions given in
Corollary \ref{CorollaryOne} are
more convenient to use in practical applications since one does not need to
check the condition (\ref{TheoremCondiiton}) for all time intervals.
Corollary \ref{CorollaryOne} tells that consensus in linear multi-agent
systems with switching directed topologies with each possible topology
containing a directed spanning tree can be achieved if the open-loop agent
dynamics are stabilizable and the dwell time is larger than a threshold
value given on the right-hand side of (\ref{TheoremCondiitonCorollary}).
\end{remark}

\begin{remark}\label{ControllableAnyDwelltime}
Suppose that $(A,\,B)$ is controllable,  it can be seen from Corollary
\ref{CorollaryOne} that the
consensus problem
  for the multi-agent system
(\ref{ClosedLoopMAs}) with an arbitrarily given dwell time $\tau_{0}$
is solved by
  the protocol
  (\ref{CommunicationProtocol}) designed in Theorem
\ref{MainTheoremOne} with an appropriately selected $\beta$.
 \end{remark}

\begin{remark}\label{RemarkOnFinalConsensusValue}
Since the underlying topology describing
  interactions between the agents is time-varying,
even though the intrinsic dynamics of each
agent are described by a linear time-invariant system, the close-loop agent
dynamics resulting from the application of
  the switching protocol proposed
in the present paper are indeed nonlinear. It is thus challenging or even
impossible to predict the final
state of consensus for such
a closed-loop multi-agent system. Clearly, the final consensus value depends on the intrinsic dynamics of each agent, the coupling strength $\alpha$, the feedback gain matrix $K$, and the switching mode among different topologies.
\end{remark}

\section{Simulation Example}
\begin{figure}[!t]
\centering
\includegraphics[width=2.6in]{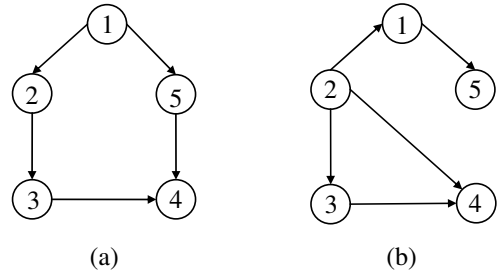}
 \caption{Communication topologies $\mathcal{G}(\mathcal{A}^{(1)})$ and $\mathcal{G}(\mathcal{A}^{(2)})$. }
\label{FigureOne}
\end{figure}

Consider a multi-agent system (\ref{ClosedLoopMAs}) consisting of six
agents, whose topology switches
between the graphs
$\mathcal{G}(\mathcal{A}^{(1)})$ and $\mathcal{G}(\mathcal{A}^{(2)})$ shown
in Figs. \ref{FigureOne}(a) and \ref{FigureOne}(b), respectively.
For convenience,  the weights on each edge are set to be $1$.
Each agent represents a vertical take-off and landing (VTOL) aircraft.
According to~\cite{FeiJi},
the dynamics of the $i$th VTOL aircraft for a typical loading and flight
condition at the air speed of $135kt$ can be described by the system
(\ref{Mode}), with
$x_{i}(t)=(x_{i1}(t),x_{i2}(t),x_{i3}(t),x_{i4}(t))^{T}\in \mathbf{R}^{4}$,
\begin{eqnarray*}
A=\left(\begin{array}{cccc}
-0.0366& 0.0271 &0.0188& -0.4555\\
0.0482&-1.01& 0.0024& -4.0208\\
0.1002 &0.3681& -0.707&1.420\\
0.0&0.0&1.0&0.0
\end{array}\right),
\end{eqnarray*}
\begin{eqnarray*}
B=\left(\begin{array}{cc}
0.4422&0.1761\\
3.5446&-7.5922\\
-5.52&4.49\\
0.0&0.0
\end{array}\right),
\end{eqnarray*}
where the state variables are defined as: $x_{i1}(t)$ is the horizontal velocity,
$x_{i2}(t)$ is the vertical velocity,
  $x_{i3}(t)$ is the pitch rate, and
$x_{i4}(t)$ is the
pitch angle \cite{FeiJi}. It can be seen from Fig. 1(a) that
$\mathcal{G}(\mathcal{A}^{(1)})$ contains
a directed spanning tree with node
$1$ as the leader, while Fig. 1(b) shows that
$\mathcal{G}(\mathcal{A}^{(2)})$ contains
a directed spanning tree rooted at
node $2$. This reflect a hypothetical situation where the communications
between aircraft 5 and 4 is unreliable, and to compensate for this, the
network switches between the two topologies, with aircraft 2 assuming the role
of the formation leader during the time intervals when it supplies
information to agent 4.

The transformed Laplacian matrices $\widehat{\mathcal{L}}^{(1)}$,
$\widehat{\mathcal{L}}^{(2)}$ in this example are
\begin{eqnarray*}
\small{\widehat{\mathcal{L}}^{(1)}{=}\left(\!\!\begin{array}{cccc}
     1 &    0 &    0 &    0\\
     0  &   1  &   0  &   0\\
     1  &  -1  &   1  &   0\\
     1  &   0  &  -1  &   2
\end{array}\!\!\right),\quad \widehat{\mathcal{L}}^{(2)}{=}\left(\!\!\begin{array}{cccc}
     2  &  -1  &   0  &   0\\
     1 &    0 &    0   &  0\\
     1  &  -1  &   1  &   0\\
     1 &   -1 &   -1 &    2
\end{array}\!\!\right)}.
\end{eqnarray*}

\begin{figure}[!t]
\centering
\includegraphics[width=2.8in]{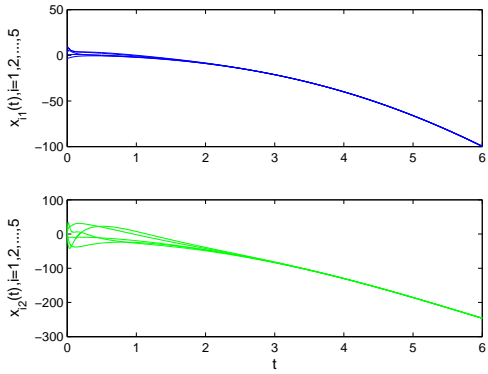}
 \caption{State trajectories $x_{i1}(t)$ and $x_{i2}(t)$, $i=1,2,\cdots,5$. }
\label{FigureStateOne}
\end{figure}

\begin{figure}[!t]
\centering
\includegraphics[width=2.8in]{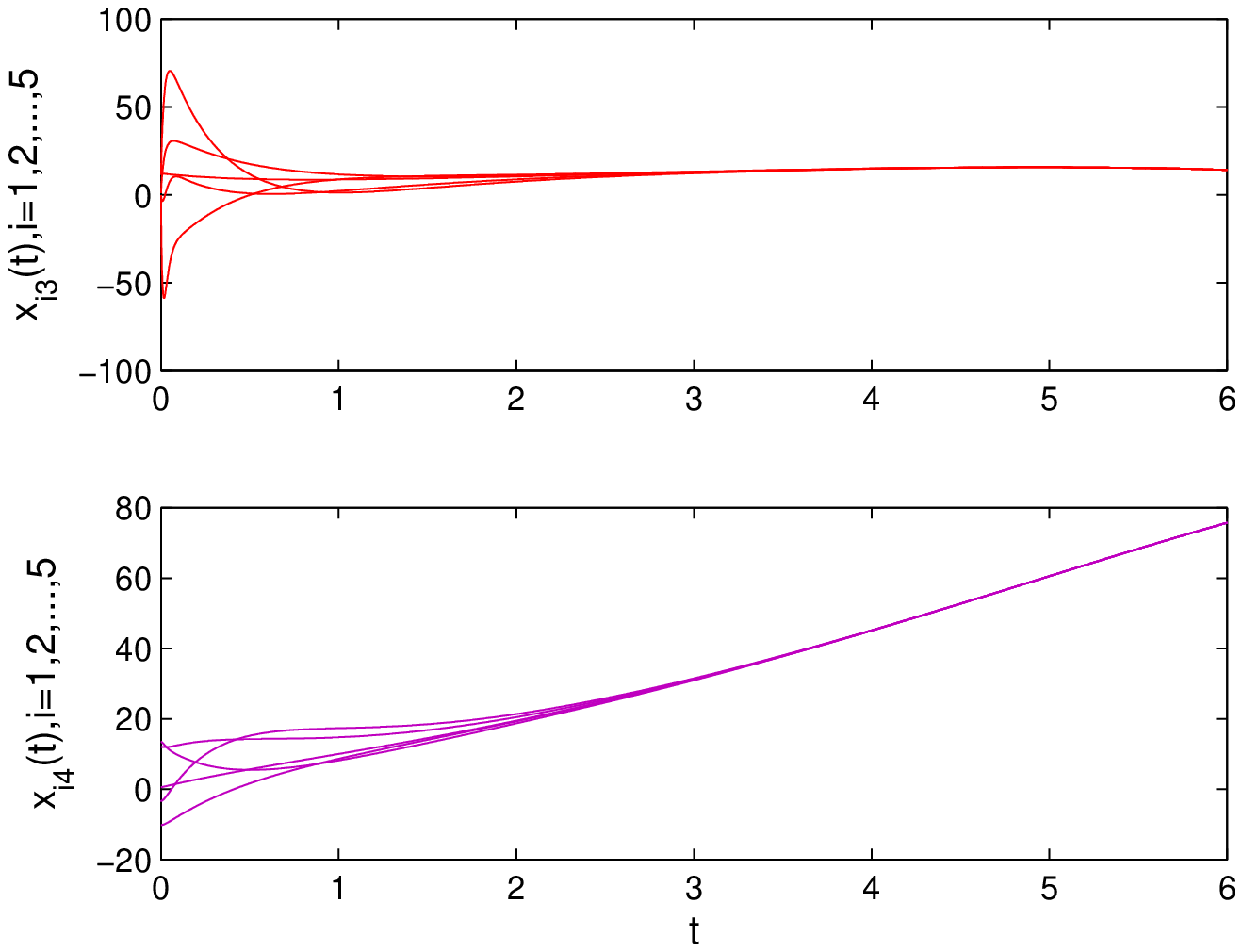}
 \caption{State trajectories $x_{i3}(t)$ and $x_{i4}(t)$, $i=1,2,\cdots,5$. }
\label{FigureStateTwo}
\end{figure}

\begin{figure}[!t]
\centering
\includegraphics[width=2.8in]{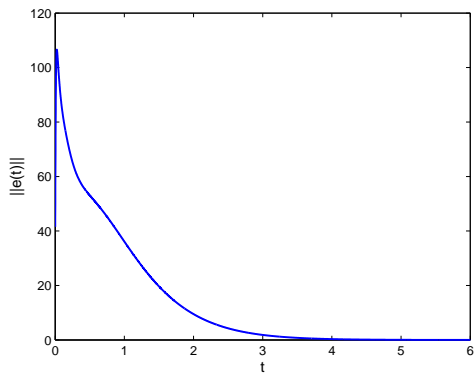}
 \caption{Evolution of $\|e(t)\|$. }
\label{FigureConsensusError}
\end{figure}

Set $c_{1}=c_{2}=0.25$. Solving the LMI (\ref{LMILaplacianMatrix}) gives
that $\overline{\lambda}_{\mathrm{max}}=3.3225$, where
$\overline{\lambda}_{\mathrm{max}}$ is defined in Corollary
\ref{CorollaryOne}. Let $\beta=3$, solving LMI
(\ref{LMISolvingFeedbackgainMatrixCorollary}) gives
that $$K=\left(\begin{array}{cccc}
 5.8206&    0.2978 &  -0.2615  & -2.7967\\
   -1.1646 &  -0.4522 &   0.0530  &  2.0420
\end{array}\right).$$  Set $\alpha=8.1>2/c_{0}=8.0$. Then, according to
Corollary \ref{CorollaryOne}, one knows that consensus in the closed-loop
multi-agent system (\ref{ClosedLoopMAs}) can be achieved if the dwell time
$\tau_{0}>  0.4002$. In simulations, let the topology switches among graph
$\mathcal{G}(\mathcal{A}^{(1)})$ and $\mathcal{G}(\mathcal{A}^{(2)})$ every
$0.5$ second. The state trajectories of the closed-loop multi-agent system
(\ref{ClosedLoopMAs}) are shown in Figs. \ref{FigureStateOne},
\ref{FigureStateTwo}. The evolution of $\|e(t)\|$ is shown in
Fig. \ref{FigureConsensusError}, which
confirms that the multi-agent system (\ref{ClosedLoopMAs}) achieves
consensus.

\section{CONCLUSIONS}
In this paper, distributed consensus of linear
multi-agent systems has been studied where the underlying topology of agent
interconnections is assumed to be time-varying. The commonly used assumption
in the existing literature that the open loop agent dynamics
are Lyapunov stable has been removed in the present work. Under a mild
condition that each possible topology only contains a directed spanning
tree and the open loop agent dynamics are stabilizable, sufficient
conditions for consensus, in terms of the graph dwell time, have been
derived and discussed. Future work will
focus on solving the distributed consensus of linear or nonlinear
multi-agent systems with switching directed topologies where some of the
possible topologies do not contain a directed spanning tree.

\addtolength{\textheight}{-12cm}   % This command serves to balance the column lengths

%%%%%%%%%%%%%%%%%%%%%%%%%%%%%%%%%%%%%%%%%%%%%%%%%%%%%%%%%%%%%%%%%%%%%%%%%%%%%%%%

%%%%%%%%%%%%%%%%%%%%%%%%%%%%%%%%%%%%%%%%%%%%%%%%%%%%%%%%%%%%%%%%%%%%%%%%%%%%%%%%


\begin{thebibliography}{99}
\bibitem{SaberFaxMurray2007Survey}
R. Olfati-Saber, J. A. Fax, and R. M. Murray, Consensus and cooperation
in networked multi-agent systems, \emph{Proc. IEEE}, vol. 95, no. 1, pp. 215-233, 2007.

\bibitem{RenBeardAtkins2007Survey}
W. Ren, R. W. Beard, and E. Atkins,
Information consensus in multivehicle cooperative control: Collective group behavior through local interaction,
\emph{IEEE Control Systems Magazine}, vol. 27, no. 2, pp. 71-82, 2007.

\bibitem{Ugrinovskii2011Automatica}
V. Ugrinovskii,
Distributed robust filtering with $H_{\infty}$ consensus of
estimates,
\emph{Automatica}, vol. 47, no. 1, pp. 1-13, 2011.

\bibitem{MeiZhangCaoBook2011}
S. Mei, X. Zhang, and M. Cao. \emph{Power Grid Complexity.} Springer-Verlage: Berlin. 2011.

\bibitem{Jadbabaie2003TAC}
A. Jadbabaie, J. Lin, and A.S. Morse, Coordination of groups
of mobile autonomous agents using nearest neighbor rules,
\emph{IEEE Transactions on Automatic Control}, vol. 48, no. 6,
pp. 988-1001, 2003.

\bibitem{SaberMurrayTAC2005}
R. Olfati-Saber and R. M. Murray,
Consensus problems in networks of agents with
switching topology and time-delays,
\emph{IEEE Transactions on Automatic Control}, vol. 49, no. 9,
pp. 988-1001, 2004.



\bibitem{RenBeard2005TAC}
W. Ren and R. W. Beard, Consensus seeking in multiagent
systems under dynamically changing interaction topologies,
\emph{IEEE Transactions on Automatic Control}, vol. 50, no. 5,
pp. 655-661, 2005.

\bibitem{Moreau2005TAC}
L. Moreau,
Stability of multiagent systems with time-dependent communication links,
\emph{IEEE Transactions on Automatic Control}, vol. 50, no. 2,
pp. 169-182, 2005.

\bibitem{RenJIRS2007}
W. Ren,
On constrained nonlinear tracking control of a small fixed-wing UAV,
 \emph{Journal of Intelligent and Robotic Systems}, vol. 48, no. 4,  pp. 525-537, 2007.

\bibitem{RenAtkinsRNC2007}
W. Ren and E. Atkins,
Distributed multi-vehicle coordinated control
via local information exchange,
\emph{Int. J. Robust and Nonlinear Control},
vol. 17, nos. 10-11, pp. 1002-1033, 2007.

\bibitem{WenDuanYuChenRNC2012}
G. Wen, Z. Duan, W. Yu, and G. Chen,
Consensus in multi-agent
systems with communication constraints,
\emph{Int. J. Robust and Nonlinear Control}, vol. 22, no. 2, pp. 170-182, 2012.

\bibitem{WielandKimFrankIJSS2011}
P. Wieland, J. Kim, and F. Allg\"{o}wer,
On topology and dynamics of consensus among linear high-order agents,
\emph{International Journal of Systems Science}, vol. 42, no. 10, pp. 1831-1842, 2011.


\bibitem{LiDuanChenHuangTCASI2010}
Z. Li, Z. Duan, G. Chen, and L. Huang,
Consensus of multiagent systems
and synchronization of complex networks: a unified viewpoint,
\emph{IEEE Transactions on Circuits Systems I: Regular Papers},
vol. 57, no. 1, pp. 213-224, 2010.

\bibitem{ZhangFrankDas2011TAC}
H. Zhang, F. L. Lewis, and A. Das,
Optimal design for synchronization of cooperative
systems: State feedback, observer and output feedback,
\emph{IEEE Transactions on Automatic Control}, vol. 56, no. 8, pp. 1498-1952, 2011.

\bibitem{TrentelmanTakabaMonshizadehIEEETAC2013}
H. L. Trentelman, K. Takaba, and N. Monshizadeh,
Robust synchronization
of uncertain linear multi-agent systems,
 \emph{IEEE Transactions on Automatic Control}, vol. 58, no. 6, pp. 1511-1523, 2013.


\bibitem{SuHuangTAC2012}
Y. Su and J. Huang
Stability of a class of linear switching systems with applications to two consensus problems,
\emph{IEEE Transactions on Automatic Control}, vol. 57, no. 6, pp. 1420-1430, 2012.

\bibitem{WuQinYuAllgowerCDC2013}
J. Wu, J. Qin, C. Yu, and F. Allg\"{o}wer,
Leaderless synchronization of linear multi-agent systems under
directed switching topologies: An invariance approach,
\emph{Proc. 52nd IEEE Conf. Decision and Control}, pp. 4043-6048, 2013.

\bibitem{WenHuYuChenCaoSCL2013}
G. Wen, G. Hu, W. Yu, J. Cao, and G. Chen,
Consensus tracking for higher-order multi-agent systems with switching directed topologies and occasionally missing control inputs,
\emph{Systems \& Control Letters}, vol. 62, no. 12, pp. 1151-1158, 2013.

\bibitem{WenDuanRenChenRNC}
G. Wen, Z. Duan, W. Ren, and G. Chen,
Distributed consensus of multi-agent systems with general linear
node dynamics and intermittent communications,
\emph{Int. J. Robust and Nonlinear Control}, doi: 10.1002/rnc.3001.

\bibitem{MinyiHuangTAC}
M. Huang and J. H. Manton, Stochastic consensus seeking with noisy and directed
inter-agent communication: Fixed and
randomly varying topologies,
\emph{IEEE Transactions on Automatic Control}, vol. 55, no. 1, pp. 235-241, 2010.

\bibitem{WenwuYuSMCB}
W. Yu, G. Chen, M. Cao, and J. Kurths,
Second-order consensus for multiagent systems with directed topologies and nonlinear dynamics,
\emph{IEEE Transactions on Systems, Man, and Cybernetics, Part B: Cybernetics,}
vol. 40, no. 3, pp. 881-891, 2010.

\bibitem{RenBeard2008Book}
W. Ren and R. W. Beard.
\emph{Distributed Consensus in
Multi-vehicle Cooperative Control: Theory and Applications.}
Springer-Verlag, London, 2008.


\bibitem{GaoZhuChenZhangMPE2013}
L. Gao, X. Zhu, W. Chen, and H. Zhang,
Leader-following consensus of linear multiagent systems with state observer under switching topologies,
\emph{Mathematical Problems in Engineering}, vol. 2013, Article ID 873140, pp. 1-12, 2013.

\bibitem{DianelBook2003}
D. Liberzon,
\emph{Switching in Systems and Control.}
Birkh\"{a}user, Boston, MA, 2003.

\bibitem{HornMatrix}
R. A. Horn, C. R. Johnson. \emph{Matrix Analysis.} Cambridge University Press, Cambridge, 1990.

\bibitem{WenYuCaoHuChenASCC2013}
G. Wen, W. Yu, J. Cao,  G. Hu, and G. Chen,
Consensus control of switching directed networks with general linear node dynamics,
\emph{Proc. 9th Asian Control Conference}, pp. 1-6, 2013.

\bibitem{WenwuCaoLuSIAM2007}
W. Yu, J. Cao, and J. L\"{u},
Global synchronization of linearly hybrid coupled networks with time-varying
delay,
\emph{SIAM J. Applied Dynamical Systems,} vol. 7, no. 1, pp. 108-133, 2008.

\bibitem{SunWangXieTAC2008}
Y. Sun and L. Wang,
Consensus of multi-agent systems in directed networks
with nonuniform time-varying delays,
\emph{IEEE Transactions on Automatic Control}, vol. 54, no. 7, pp. 1607-1613, 2009.


\bibitem{FeiJi}
K. Narendra and S. Tripathi, Identification and optimization of aircraft dynamics,
\emph{Journal of Aircraft,}
vol. 10, no. 4, pp. 193-199, 1973.

%\bibitem{ZhongkuiLiIET2012}
%Z. Li, X. Liu, M. Fu, and L. Xie,
%``Global $H_{\infty}$ consensus of multi-agent systems with Lipschitz non-linear dynamics,
%\emph{IET Control Theory \& Applications}, vol. 6, no. 13, 2012.


\end{thebibliography}
\end{document}